\documentclass[twocolumn,aps,pra,superscriptaddress]{revtex4-1}
\usepackage{graphicx}
\usepackage{amsmath}
\usepackage{amssymb}
\usepackage{braket}
\usepackage{bm}
\usepackage[colorlinks=true,linkcolor=blue,citecolor=blue,urlcolor=blue]{hyperref}

\begin{document}

\title{Confinement-Induced Resonance with Weak Background Interaction}

\author{Ren Zhang}
\email{rine.zhang@gmail.com}

\affiliation{School of Science, Xi'an Jiaotong University, Shaanxi, 710049, China}

\author{Peng Zhang}
\email{pengzhang@ruc.edu.cn}

\affiliation{Department of Physics, Renmin University of China, Beijing, 100872,
China}

\affiliation{Beijing Computational Science Research Center, Beijing, 100084, China}

\date{\today}

\begin{abstract}
We studied the scattering problem of two distinguishable atoms with unequal mass, where one atom (atom $\alpha$) is trapped in a quasi-one-dimensional (quasi-1D) tube and the other one (atom $\beta$) is localized by a 3D harmonic trap. We show that in such a system if atom $\alpha$ is much heavier than $\beta$, confinement-induced resonance (CIR) can appear when the 3D $s$-wave scattering length $a_s$ of these two atoms is much smaller than the characteristic lengths (CLs) of the confinements, for either $a_s>0$ or $a_s<0$. This is quite different from the usual CIRs which occurs only when $a_s$ is comparable with the CL of confinement. Moreover, the CIRs we find are broad enough that can serve as a tool for the control of effective inter-atomic interaction. We further show the mechanism of these CIRs via the Born-Oppenheimer approximation. Our results can be used for the realization of strongly-interacting systems with ultracold atoms with weak 3D background interaction (i.e., small $a_s$), e.g., the realization of ultracold gases with strong spin-dependent interaction at zero magnetic fields.
\end{abstract}
\maketitle

\section{Introduction}

One important advantage of the quantum simulation
with ultracold atomic gases is that in such a system the inter-atomic
interaction can be efficiently controlled. The most widely-used technique
for this control is via magnetically tunable Feshbach resonance (MFR) \cite{kohler06,Chin2010}. In addition,
the confinement-induced resonance (CIR) is also a powerful tool,
with which one can tune the interaction between trapped ultracold
atoms by changing the geometric parameters of the confinements \cite{Olshanii1998,Bergeman2003,Haller2010,2dcir,Shiguo,WeiPeng,PeterZoller,HansPeter,XL_lattice,yvan,HanbergeCIR1,HanbergeCIR2,HanbergeCIR3,HanbergeCIR4}. 

In most cases, when a CIR occurs the characteristic length (CL) of
the confinement should be comparable with the $s$-wave scattering length $a_{s}$
of the two atoms in three-dimensional (3D) free space. For instance,
as shown by M. Olshanii, the CIR condition for two atoms in a quasi-1D
tube with a CL $a_{\perp}$ is $a_{\perp}/a_{s}\approx1.4603$ \cite{Olshanii1998}. This
makes sense because a ``resonance" usually appears
when the values of several characteristic parameters of the system
are similar to each other. Nevertheless, for the optical confinements
realized in current experiments, the CLs are usually larger than $1000a_{0}$
with $a_{0}$ being the Bohr's radius, while for most kinds
of cold atoms $|a_{s}|$ is below (or about) $200a_{0}$ in the absence of magnetic
field ($B=0$) \cite{exception}. As a result, to realize a CIR one has to resort to the MFR to
enhance the value of $a_{s}$. So far the only exception is the CIR
of $^{173}$Yb atoms featured by an extremely large background scattering length
(about $2000a_{0}$ \cite{Yb173as}) \cite{kondo_exp}.

Therefore, it is natural to ask if the above
limitation can be broken or not. Namely, can a specific CIR which satisfies
the following two conditions are realized? 
\begin{itemize}
\item[(i)] The CIR can occur when the CLs of the confinements are much larger than the scattering length $|a_{s}|$. 
\item[(ii)] The CIR is broad enough so that it can be used as an efficient tool
for the control of effective interaction between ultracold atoms.
\end{itemize}

Various interesting phenomena may benefit from such a specific CIR. For instance, such a
CIR can occur at zero magnetic fields ($B=0$), for which different atomic hyperfine
states are degenerate, and thus spin-changing scattering processes
between hyperfine spin channels are energetically permitted \cite{spin-dynamics,Resonant-se}. Using
this CIR one can control these processes and then realize systems with
strong inter-atomic spin-spin interaction, e.g., the
spin-exchange interaction, which is important for the quantum simulation
of the Kondo effect or other magnetic effects \cite{kondobook,spin-exchange1,spin-exchange2,renKondo1,yantingkondo,renKondo2,kondo_exp,Takahashiyb171}. Notice that these spin-spin interactions
cannot be controlled via a usual MFR because spin-changing scattering
processes are energetically suppressed by the Zeeman-energy gap between
different hyperfine channels, which are induced by the magnetic field of the MFR \cite{yb}. Besides, such a specific CIR can be realized without the help of a magnetic field. Thus, the magnetic field can be reserved 
 for purposes
 other than the control of inter-atomic interaction, e.g., the trapping of atoms
in an atom chip \cite{atomchip}. Moreover, since this CIR can occur for small 3D scattering
length $a_{s}$, as shown below, the collisional losses may be suppressed.

In previous studies, it has been demonstrated that 
for ultracold gases with {\it negative} 3D scattering
length $a_{s}$ (i.e., $a_s<0$), a specific CIR which satisfies the
conditions (i) and (ii) can be realized in two cases, i.e., the scattering
between two atoms in a 3D isotropic square optical
lattice \cite{PeterZoller,HansPeter,XL_lattice}, and the scattering between a heavy atom freely moving in 3D space and a light atom localized in a 3D harmonic trap \cite{yvan}. Nevertheless,
 for most species of ultracold
atoms in current experiments we have $a_{s}>0$ for $B=0$. To our
knowledge, the specific CIR which can occur for both $a_{s}>0$
and $a_{s}<0$ has not been discovered before.

In this manuscript, we propose that such a specific CIR can be realized 
for a mixed-dimensional ultracold gas with either {\it positive} or {\it negative} 3D scattering
length $a_{s}$.
Explicitly, we consider the scattering between an atom (atom $\alpha$) moving in quasi-1D confinement and another atom (atom $\beta$) localized in a 3D harmonic trap, and thus behaves as a quasi-0D impurity (Fig.~\ref{schematic}). 
We find that in such a ``quasi-(1+0)D" system a CIR which satisfies
all the two conditions (i) and (ii) can occur when atom $\alpha$
is much heavier than the atom $\beta$, for either $a_s>0$ or $a_s<0$. 

We further explain the mechanism of the specific CIRs in our system with an
analysis based on the Born-Oppenheimer approximation (BOA). Explicitly, we find that
the specific CIR of our system mainly results from both of the following two facts:

(A): As specified by the BOA, the heavy atom $\alpha$ can experience a 1D finite-range potential
$V_{\rm BOA}(z_\alpha)$ along the axial direction of the quasi-1D confinement (the $z$-direction),
 which is induced by the light atom $\beta$. 
 In addition, for the case with small $|a_s|$, $V_{\rm BOA}(z_\alpha)$ is proportional to the strength of the 3D Huang-Yang pseudo potential between the atoms $\alpha$ and $\beta$, i.e., $a_s/\mu$, with $\mu$ being the reduced mass of these two atoms. For our system with $\alpha$ being much heavier than $\beta$, $\mu$ is approximately the mass of atom $\beta$. Therefore, for small $|a_s|$, the 1D potential $V_{\rm BOA}(z_\alpha)$ for atom $\alpha$ can still be strong under the condition that the atom $\beta$ is light enough.


(B) When $a_s>0$, $V_{\rm BOA}(z_\alpha)$ is a 1D finite-range purely-repulsive potential, i.e., a 1D potential barrier. Intuitively speaking, there should be no resonance in such a system, because there is no bound state. However, a low-energy scattering resonance can still appear when the height of $V_{\rm BOA}$ takes some certain values. We illustrate this effect with an analytically-solvable 1D square-barrier model. As a result of this effect, when one tunes $V_{\rm BOA}$ via the CLs of the confinements, a CIR can be induced. On the other hand, when $a_s<0$, $V_{\rm BOA}(z_\alpha)$ is a 1D attractive potential well with the depth dependent on the CLs of the confinements. Thus, it is quite natural that a CIR can be induced when one tunes the depth of $V_{\rm BOA}(z_\alpha)$ to some certain value by changing the CLs of the confinements.

Our results are helpful for realizing strong effective inter-atomic interaction of ultracold atoms in quasi-1D confinement with localized impurities, which can be either spin-dependent or spin-independent. Such a system can be used for the study of Kondo physics \cite{ReyNP,renKondo1,yantingkondo,qingjikondo,renKondo2,Kuzmenko}, quantum open system \cite{opensystem} and precision measurement \cite{BmeasureL,BmeasureA}. Also, our result shows the existence of broad CIRs for small {\it positive} 3D scattering length, and thus implies the possibility of finding such CIRs in more general systems. 

The remainder of this manuscript is organized as follows. In Sec. II we solve the quasi-(1+0)D scattering problem of two atoms with unequal mass and show the appearance of the specific CIRs. In Sec. III we explain our result with the analysis based on BOA. In particular, we illustrate that in the 1D cases a low-energy scattering resonance can be induced by a purely-repulsive potential barrier. A summary of our results is given in Sec. IV. In the appendix, we show some details of our calculation.


\section{ CIR in a quasi-(1+0)D system}

In this section, we first introduce the quasi-(1+0)D system studied in this work, and then give the definition of CIR for our system. After that, we illustrate our numerical results which show that specific CIRs satisfying the above conditions (i) and (ii)  can appear in our system, for either $a_s<0$ or $a_s>0$. 

\subsection{System and Hamiltonian}

\begin{figure}
\centering
\includegraphics[width=0.4\textwidth]{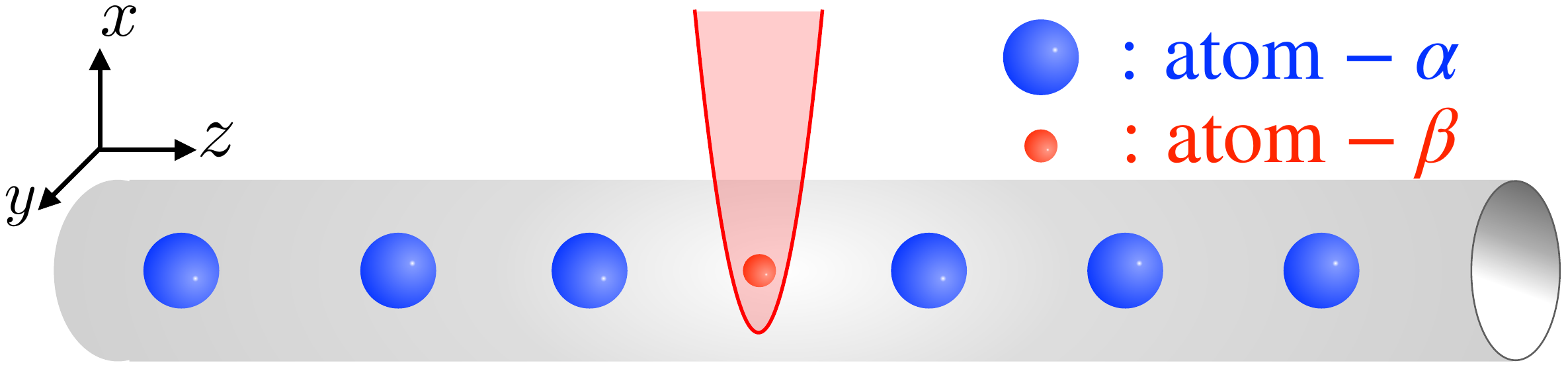}
\caption{Schematic diagram of our quasi-(1+0)D system. Atom-$\alpha$ and $\beta$ are simultaneously confined in a quasi-1D tube which is described by a 2D isotropic harmonic potential with frequency $\omega_\perp$ in the $x$-$y$ plane. Atom-$\beta$ is further trapped by an external harmonic trap along the $z$-direction, which is centered at the origin of the $z$-axis and has frequency $\omega_z$.\label{schematic}}
\end{figure}

As shown in Fig.~\ref{schematic}, we consider the two-body problem with a heavy atom $\alpha$
and a light atom $\beta$. The
 atom $\alpha$ is moving in a quasi-1D
confinement along the $z$-direction (axial direction) which is described
as a 2D isotropic harmonic potential with frequency
$\omega_{\perp}$ in the $x-y$ plane (transverse plane), and atom $\beta$ is trapped in a 3D harmonic potential with frequencies $\omega_{xy}$ and $\omega_z$ in the $x-y$ plane and $z$-direction, respectively.
For simplicity, here we assume the confinements of atoms $\alpha$ and $\beta$ have the same transverse frequencies, i.e., $\omega_{xy}=\omega_\perp$.

As a result of the above assumption $\omega_{xy}=\omega_\perp$, the transverse relative motion of these two atoms can be decoupled
from their center-of-mass motion. Thus, in the two-body problem we can ignore the center-of-mass motion in the $x$- and $y$-directions. Therefore,
the Hamiltonian for this system is 
\begin{align}
H=H_{0}+V.\label{h}
\end{align}
Here $H_{0}$ is the free Hamiltonian and can be expressed as
\begin{align}
 H_{0}=H_{\perp}+H_{z}^{(\alpha)}+H_{z}^{(\beta)},\label{2}
 \end{align}
where
\begin{align}
H_{\perp}=-\frac{\hbar^{2}}{2\mu}\nabla_{\bm{\rho}}^{2}+\frac{\mu\omega_{\perp}\rho^{2}}{2}, 
\end{align}
 is the free Hamiltonian for the transverse relative motion of the two atoms, with
 $\mu$ and ${\bm{\rho}}$ being 
the reduced mass and the relative
position of these two atoms in the $x-y$ plane, respectively. In Eq. (\ref{2}),
\begin{align}
H_{z}^{(\alpha)}=-\frac{\hbar^{2}}{2m_{\alpha}}\frac{\partial^{2}}{\partial z_{\alpha}^{2}}
 \end{align}
and 
 \begin{align}
H_{z}^{(\beta)}=-\frac{\hbar^{2}}{2m_{\beta}}\frac{\partial^{2}}{\partial z_{\beta}^{2}}+\frac{m_{\beta}\omega_{z}z_{\beta}^{2}}{2}
 \end{align}
 are the Hamiltonian for the motion of the atoms $\alpha$ and $\beta$ in the $z$-direction, respectively, where $m_{\alpha(\beta)}$  and $z_{\alpha(\beta)}$ are the mass and
$z$-coordiante of the atom $\alpha$ ($\beta$), respectively. For this system, we can define two CLs $a_{\perp}$ and $a_z$ of the confinement as follows:
\begin{align}
a_\perp=\sqrt{\frac{\hbar}{\mu\omega_\perp}};\quad a_z=\sqrt{\frac{\hbar}{m_\beta\omega_z}}.\label{CL}
\end{align} In this manuscript, we will focus on the case that
\begin{align}
\label{massimba}
m_\alpha\gg m_\beta.
\end{align}
In Eq.~(\ref{h}) $V$ is the 
interaction potential between the two atoms, and is
 modeled with the Huang-Yang pseudo potential, i.e., 
\begin{align}
\label{HY}
V({\bf r})=\frac{2\pi\hbar^{2}a_{s}}{\mu}\delta({\bf r})\frac{\partial}{\partial r}(r\cdot),
\end{align}
with $a_{s}$ being the 3D $s$-wave scattering length of these two atoms and
${\bf r}={\bm{\rho}}+(z_{\alpha}-z_{\beta}){\bf e}_{z}$ being the
3D relative position vector of the two atoms. Here ${\bf e}_{z}$
is the unit vector along the $z$-direction.

 \begin{figure*}
\centering
\includegraphics[width=0.25\textwidth]{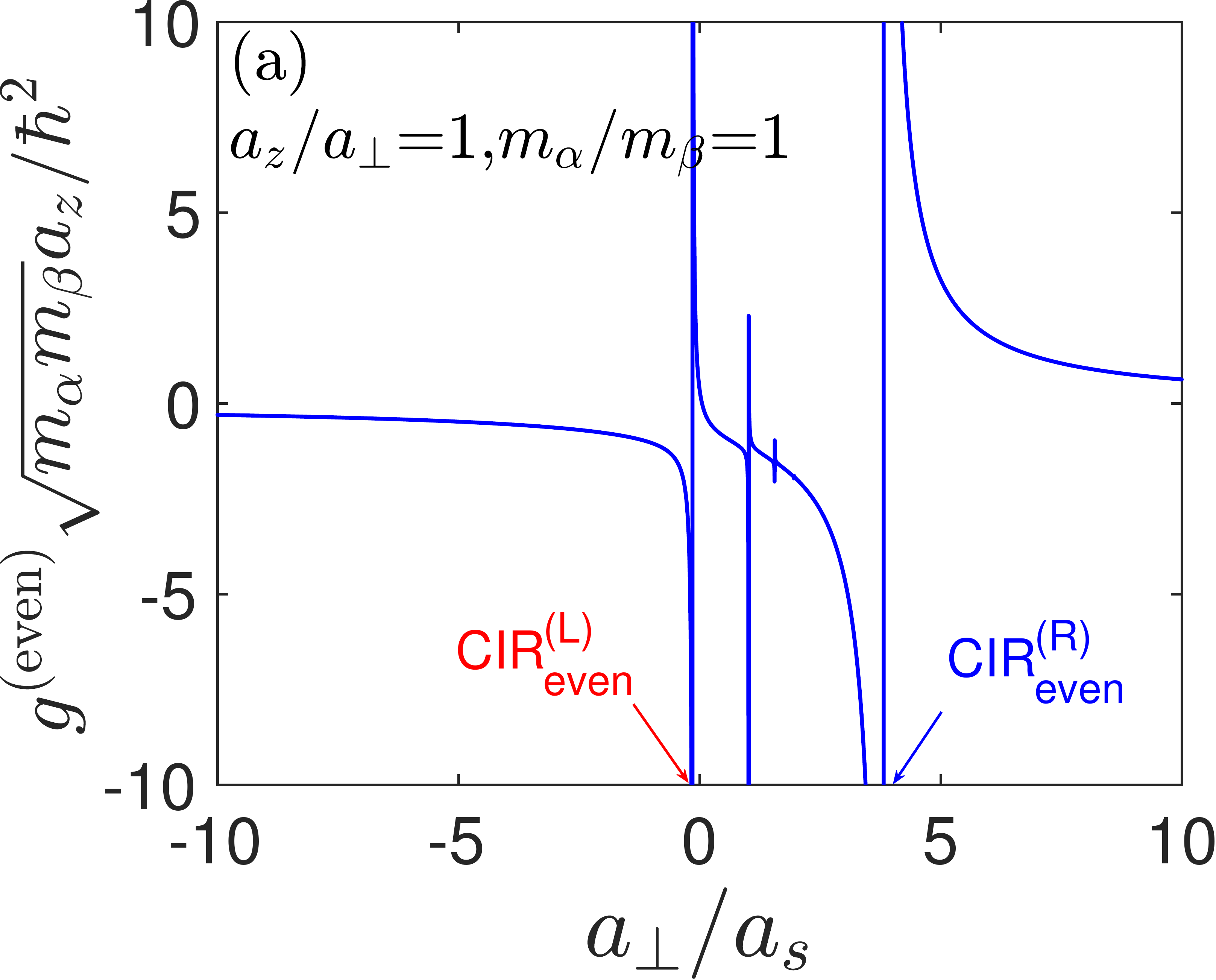}
\includegraphics[width=0.24\textwidth]{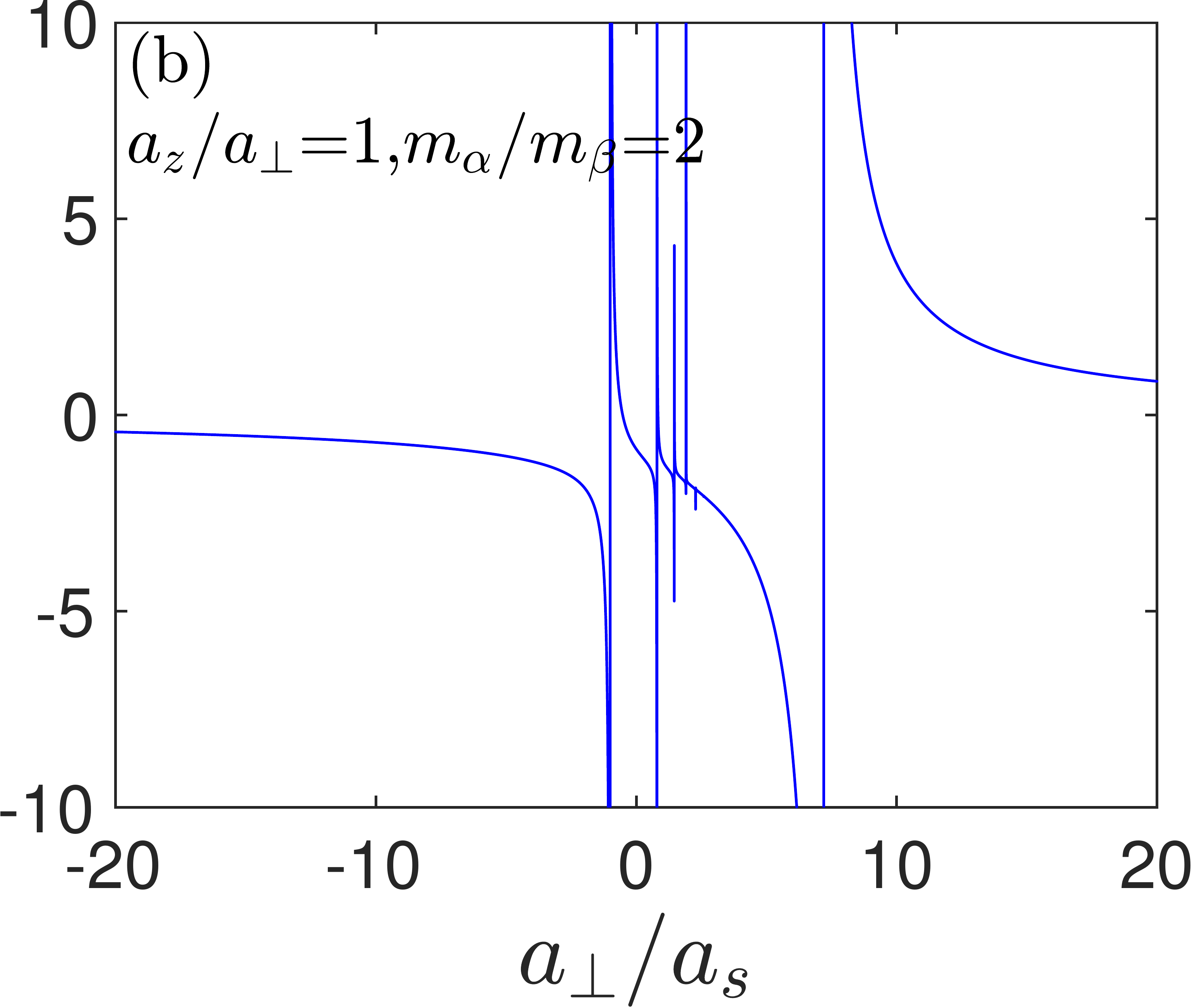}
\includegraphics[width=0.24\textwidth]{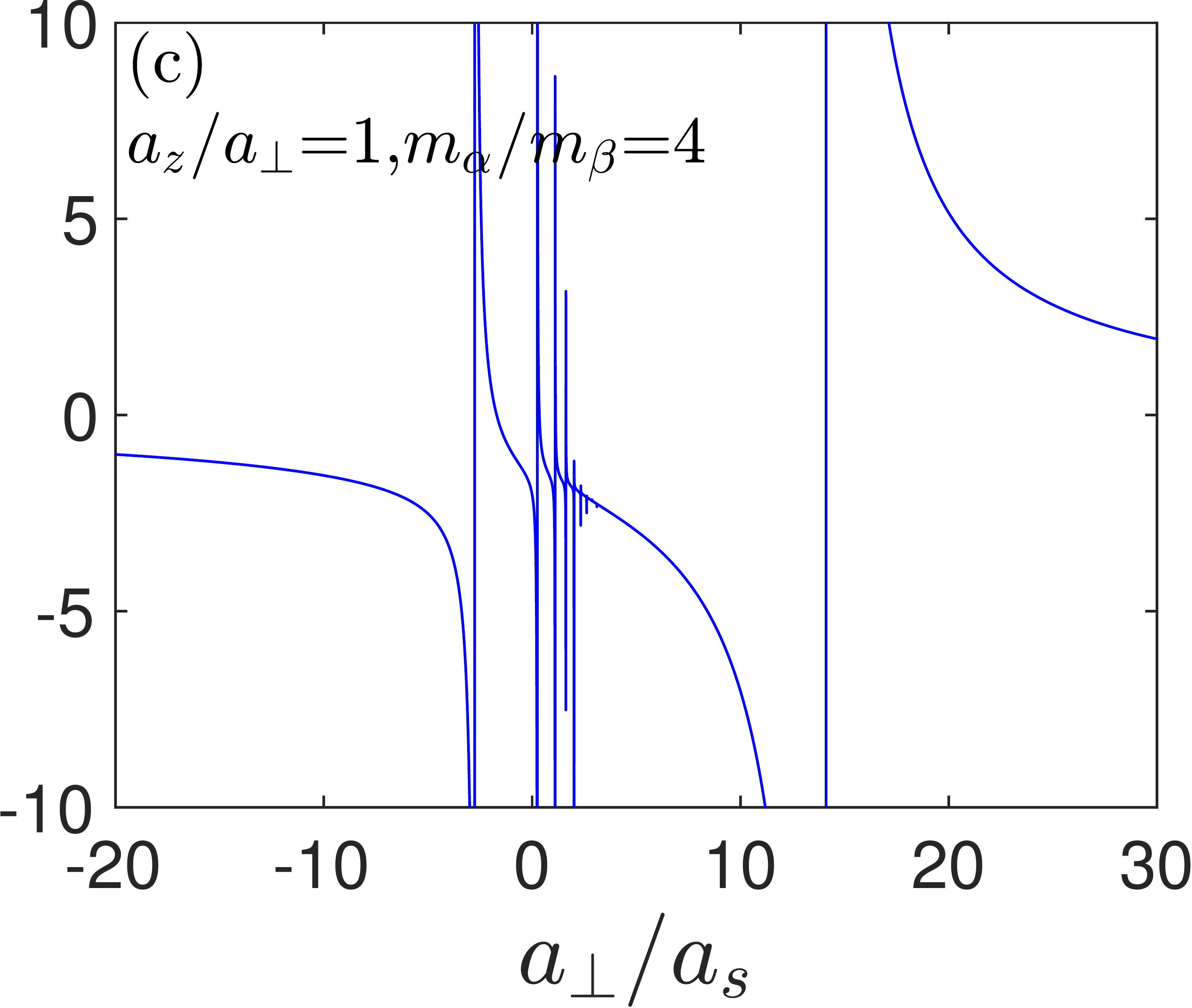}
\includegraphics[width=0.24\textwidth]{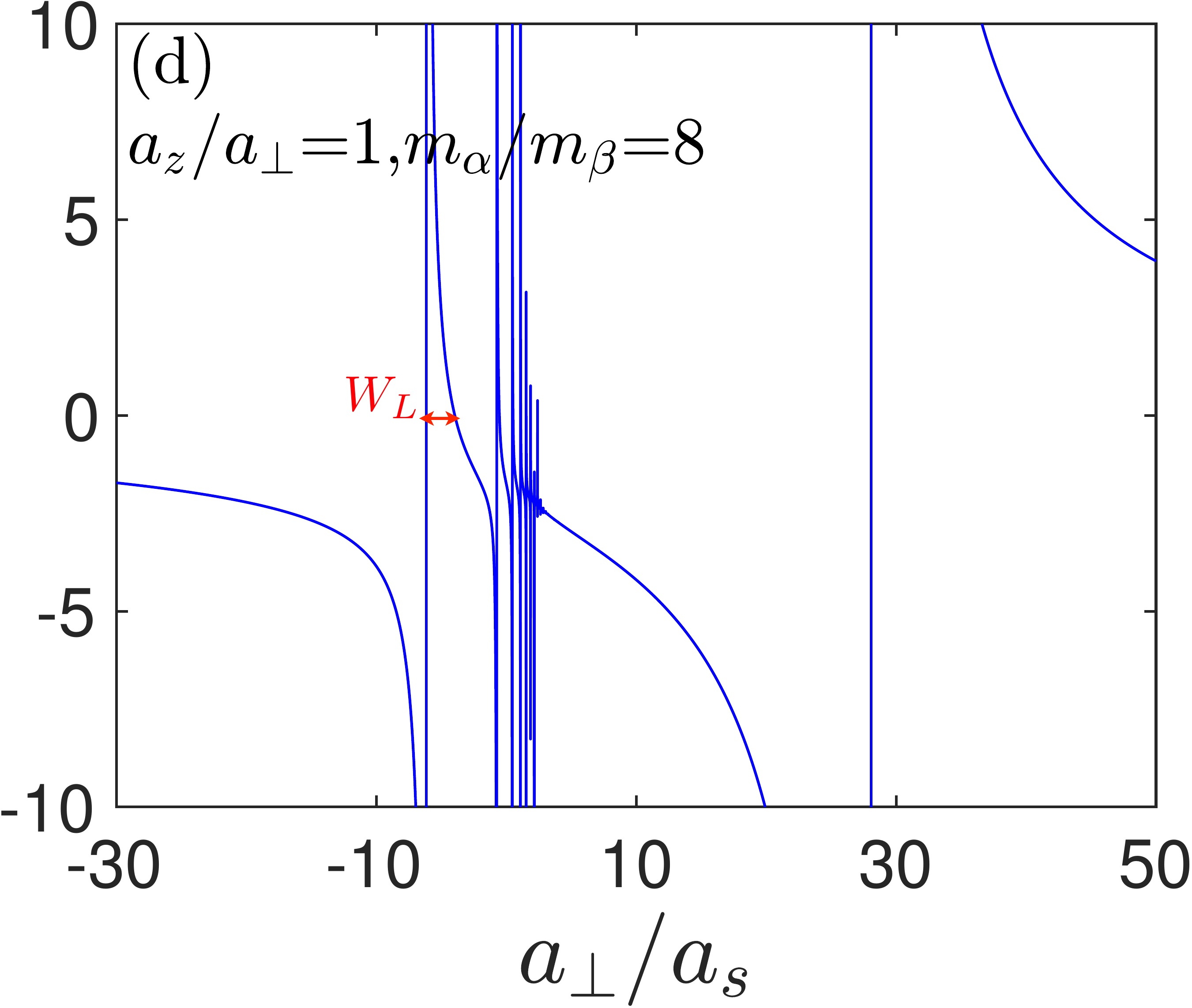}
\includegraphics[width=0.25\textwidth]{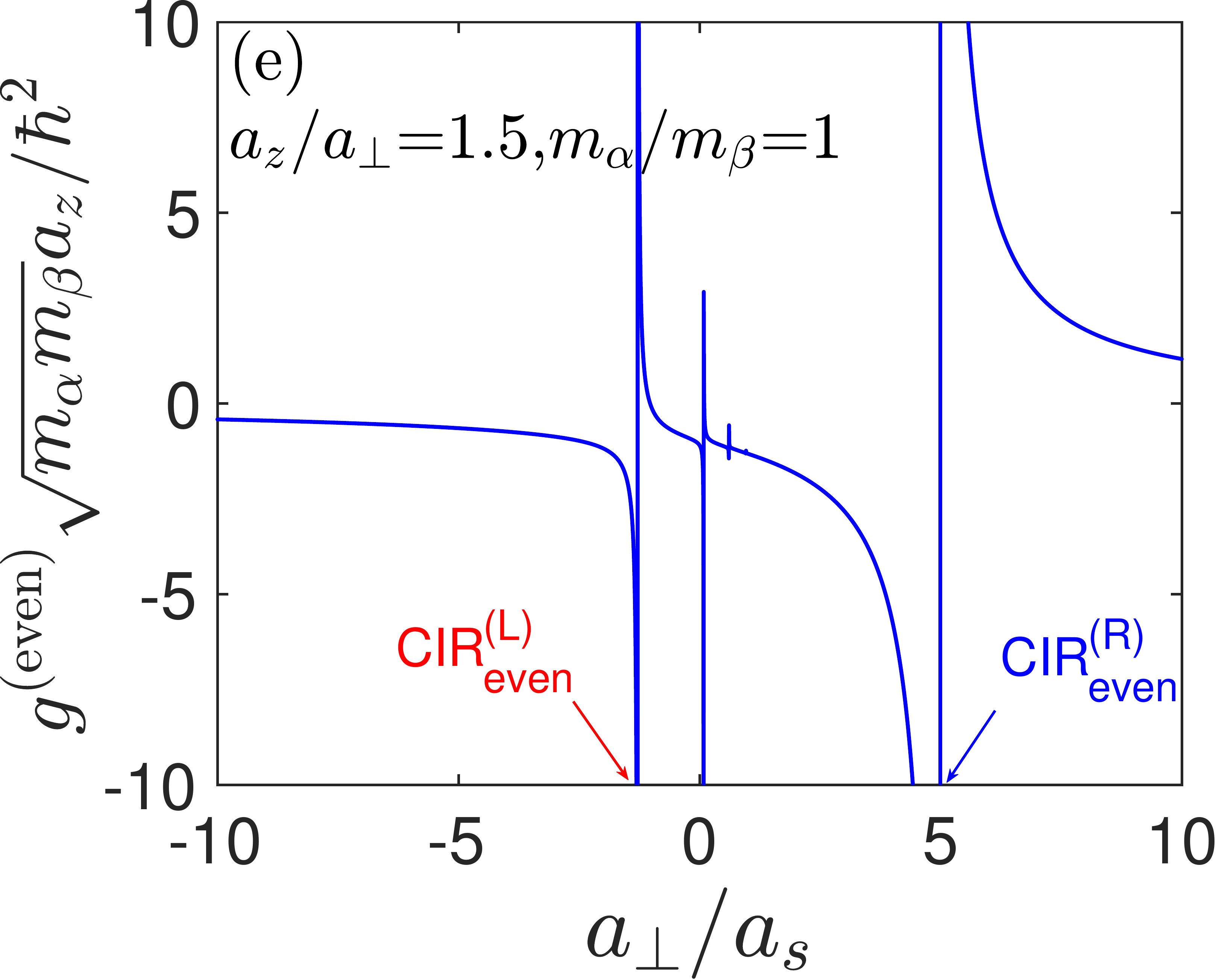}
\includegraphics[width=0.24\textwidth]{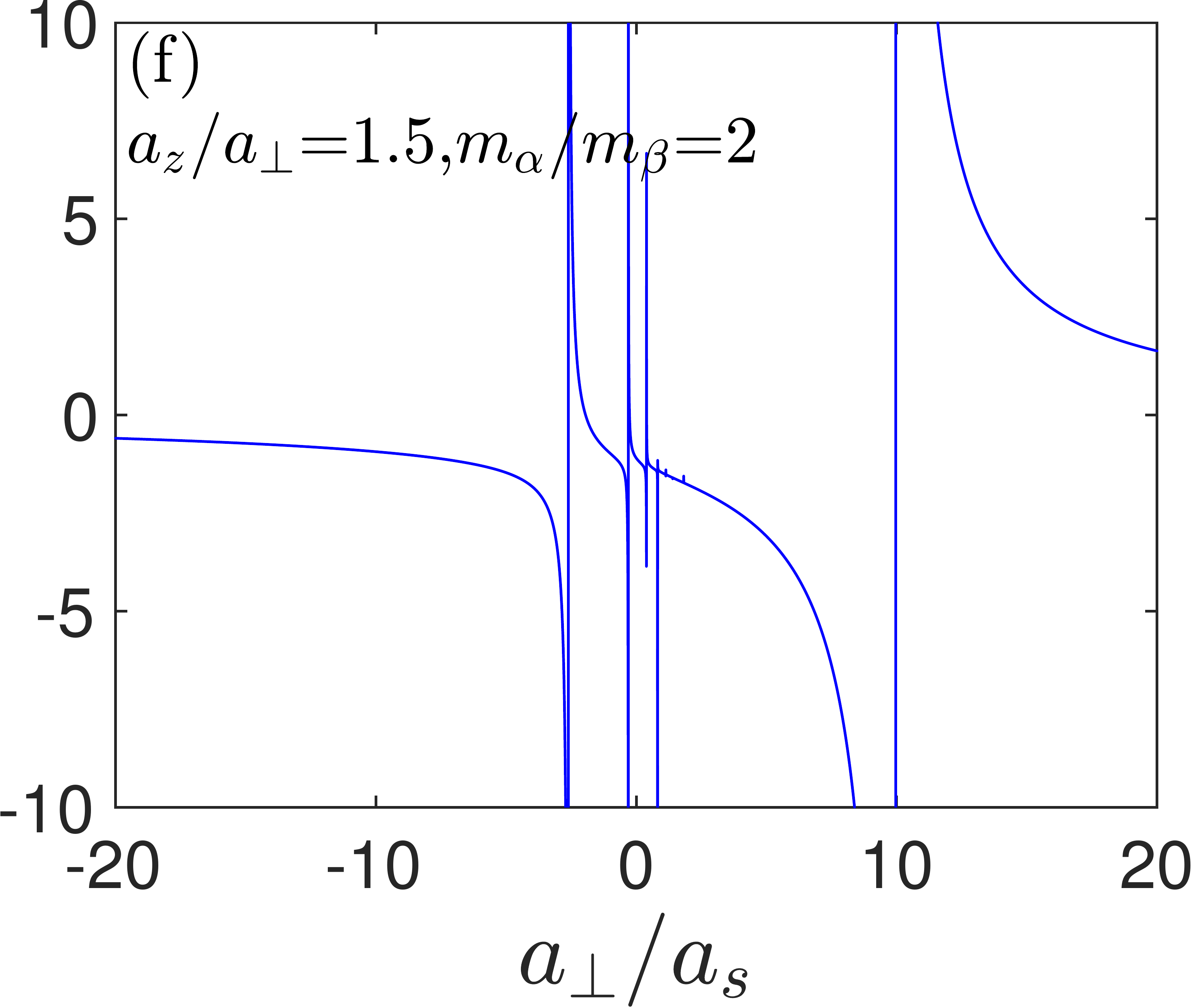}
\includegraphics[width=0.24\textwidth]{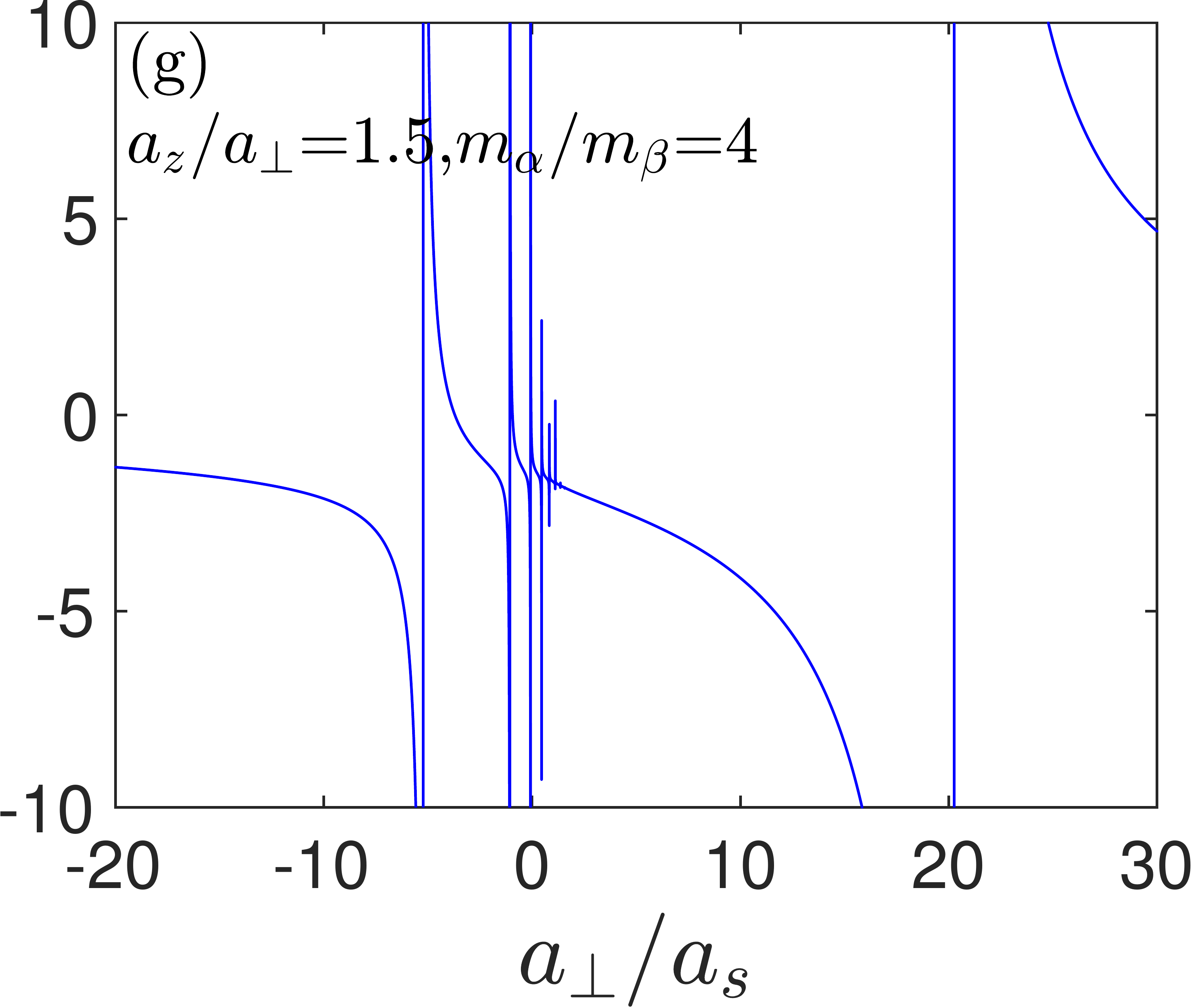}
\includegraphics[width=0.24\textwidth]{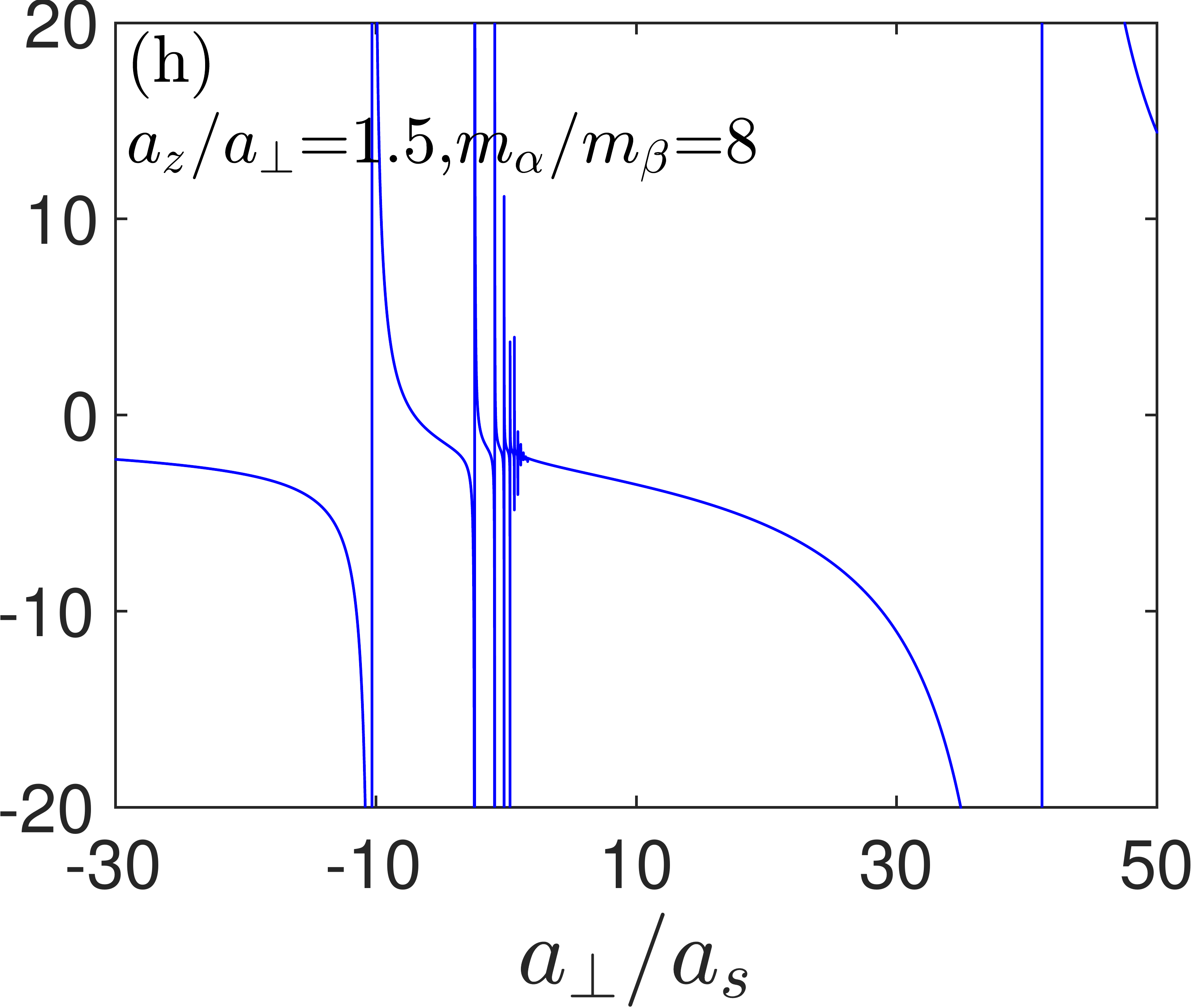}
\caption{(color online) Effective interaction strength $g^{\rm (even)}$ as a function of $a_{\perp}/a_{s}$ for different mass ratio $m_{\alpha}/m_{\beta}$ and confinement aspect ratio $a_{z}/a_{\perp}$. The two broadest CIRs located are labeled by CIR$_{\rm even}^{\rm (R)}$ and CIR$_{\rm even}^{\rm (L)}$ respectively. In (d) we further show the definition of the width $W_L$ of CIR$_{\rm even}^{\rm (L)}$.\label{fig2}}
\end{figure*}
\subsection{Low-energy scattering and CIR}

In this manuscript we consider the low-energy scattering between atoms $\alpha$ 
and $\beta$ which are in the ground state $\phi_{\perp}({\bm{\rho}})$ of $H_{\perp}$ and the
ground state $\phi_{z}(z_{\beta})$ of $H_{z}^{(\beta)}$, respectively.
Thus, the incident state for the scattering problem can be expressed
as 
 \begin{align}
 \label{incoming-state}
\Psi^{(0)}({\bm{\rho}},z_{\alpha},z_{\beta})=\frac{e^{ikz_{\alpha}}}{\sqrt{2\pi}}\phi_{z}(z_{\beta})\phi_{\perp}({\bm{\rho}}),
 \end{align}
with $k$ being the incident momentum of the atom $\alpha$. 
We consider the low-energy case where the incident kinetic energy of atom $\alpha$ is much lower than the frequencies of the transverse and axial confinements, i.e., 
\begin{align}
\label{lowenergy}
\frac{\hbar^{2}k^{2}}{2m_{\alpha}}\ll\hbar\omega_z,\hbar\omega_\perp.
\end{align}
Here we emphasis that using Eq.(\ref{CL}) we can re-express this condition as $
ka_{\perp},ka_{z}\ll\sqrt{2m_{\alpha}/m_{\beta}}.
$ Thus, when the mass ratio between the atom $\alpha$ and $\beta$ is large enough, this condition can be satisfied even when $k$ is larger than $1/a_\perp$ or $1/a_z$. Under such a low-energy condition, the corresponding
scattering state $\Psi^{(+)}({\bm{\rho}},z_{\alpha},z_{\beta})$ is
determined by the Schr$\ddot{{\rm o}}$dinger equation 
 \begin{align}
 \label{scattering-state}
H\Psi^{(+)}({\bm \rho},z_\alpha,z_\beta)=\left(\frac{\hbar^{2}k^{2}}{2m_{\alpha}}+\frac{\hbar\omega_{z}}{2}+\hbar\omega_{\perp}\right)\Psi^{(+)}({\bm \rho},z_\alpha,z_\beta)\nonumber\\
 \end{align}
and the out-going boundary condition 
\begin{align}
\label{outgoing}
 &\lim_{|z_\alpha|\rightarrow\infty}\Psi^{(+)}({\bm \rho},z_\alpha,z_\beta)\nonumber\\
 &=\frac{1}{\sqrt{2\pi}}\left\{e^{ikz_{\alpha}}+f^{{\rm (e)}}(k)e^{ik|{z_{\alpha}}|}\right.\nonumber \\
 &+\left.f^{{\rm (o)}}(k){\rm sign}[z_{\alpha}]e^{ik|{z_{\alpha}}|}\right\}\phi_{z}(z_{\beta})\phi_{\perp}({\bm{\rho}}),
\end{align}
where $f^{{\rm (e)}}(k)$ and $f^{{\rm (o)}}(k)$ are the effective 1D scattering amplitudes for the even and odd partial waves, respectively, and can be expressed as 
\begin{align}
f^{\rm (e)}(k)&\approx\frac{-1}{1+ika_{\rm e}};\label{fe}\\
f^{\rm (o)}(k)&\approx\frac{-ik}{ik+a_{\rm o}^{-1}},\label{fo}
\end{align}
in the low-energy limit. Here $a_{\rm e}$ and $a_{\rm o}$ are the even and odd wave 1D scattering lengths, respectively.
They are functions of the 3D scattering length $a_s$ as well as the CLs $a_{\perp}$ and $a_z$ of the confinement.
In other words, we have
\begin{align}
a_{\rm e/o}=a_{\rm e/o}(a_s,a_\perp,a_z).
\end{align}

In the low-energy case, when the mean value of the inter-atomic distance is much larger than the width of the matter-wave packets of these two
atoms, the transverse motion of the two atoms and the axial motion of atom $\beta$ are frozen in the ground state of the corresponding confinements. 
As a result, our system can be described by a simple pure-1D effective model for the axial motion of atom $\alpha$, with the effective Hamiltonian \cite{Girardeau}
\begin{eqnarray}
H^{\rm (eff)}=-\frac{\hbar^{2}}{2m_{\alpha}}\frac{\partial^{2}}{\partial z_{\alpha}^{2}}+V_{\rm eff}(z_\alpha),\label{heff2}
\end{eqnarray}
with
\begin{eqnarray}
V_{\rm eff}(z_\alpha)
=g^{\rm(even)}\delta(z_\alpha){\hat d}_{\rm
 e}+g^{\rm(odd)}\delta^\prime(z_\alpha){\hat d}_{\rm o}.\label{Veff}
\end{eqnarray}
Here
 $\delta(z_\alpha)$ is the Dirac delta function,
 $\delta^\prime(z_\alpha)={\frac {d}{dz_\alpha} \delta(z_\alpha)}$, and the operators
 ${\hat d}_{e}$ and ${\hat d}_{o}$ 
 are defined as
\begin{eqnarray}
{\hat d}_{\rm e}\psi(z_\alpha)&\equiv&\frac{1}{2}\left[\left.\psi(z_\alpha)\right|_{z_\alpha=0^+}+\left.\psi(z_\alpha)\right|_{z_\alpha=0^-}\right],\nonumber\\
{\hat d}_{\rm o}\psi(z_\alpha)&\equiv&\frac{1}{2}\left[\left.\frac{d}{dz_\alpha}\psi(z_\alpha)\right|_{z_\alpha=0^+}+\left.\frac{d}{dz_\alpha}\psi(z_\alpha)\right|_{z_\alpha=0^-}\right].\nonumber\\
\nonumber
\end{eqnarray}
In Eq. (\ref{Veff}) the strengths $g^{\rm(even)}$ and $g^{\rm(odd)}$ can be expressed as functions of the scattering lengths $a_{\rm e,o}$ as
\begin{eqnarray}
g^{\rm(even)}=-\frac{\hbar^2}{m_\alpha a_{\rm e}};\ \ \ g^{\rm(odd)}=\frac{\hbar^2 }{m_\alpha}a_{\rm o},\label{geven}
 \end{eqnarray}
respectively \cite{Girardeau}.
A straightforward calculation shows that the effective potential $V_{\rm eff}(z_\alpha)$ can reproduce the low-energy scattering amplitudes $f^{\rm (e,o)}(k)$ given by the exact ``quasi-1D+quasi-0D" Hamiltonian $H$. 
\begin{figure}[h]
\centering
\includegraphics[width=0.38\textwidth]{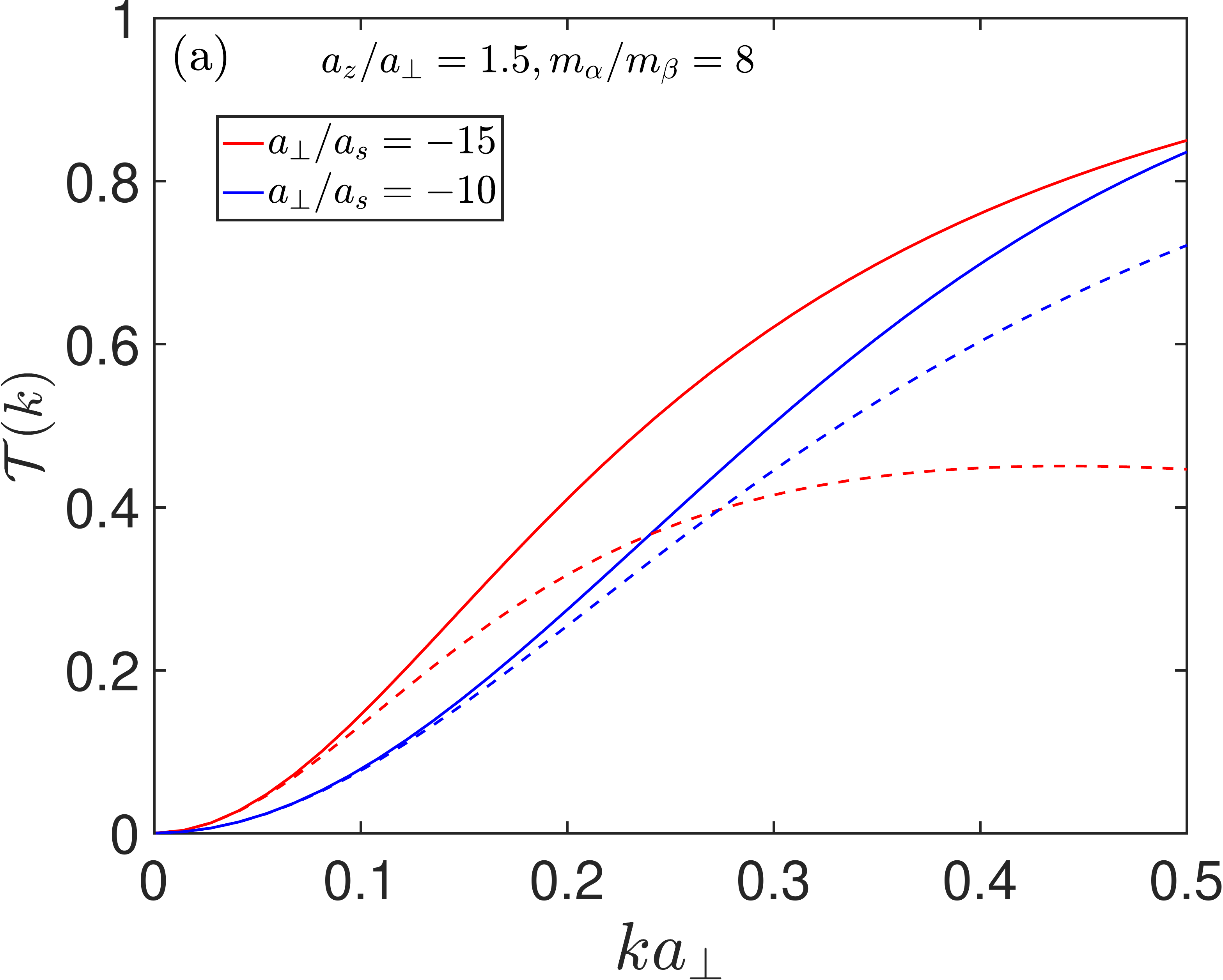}
\includegraphics[width=0.38\textwidth]{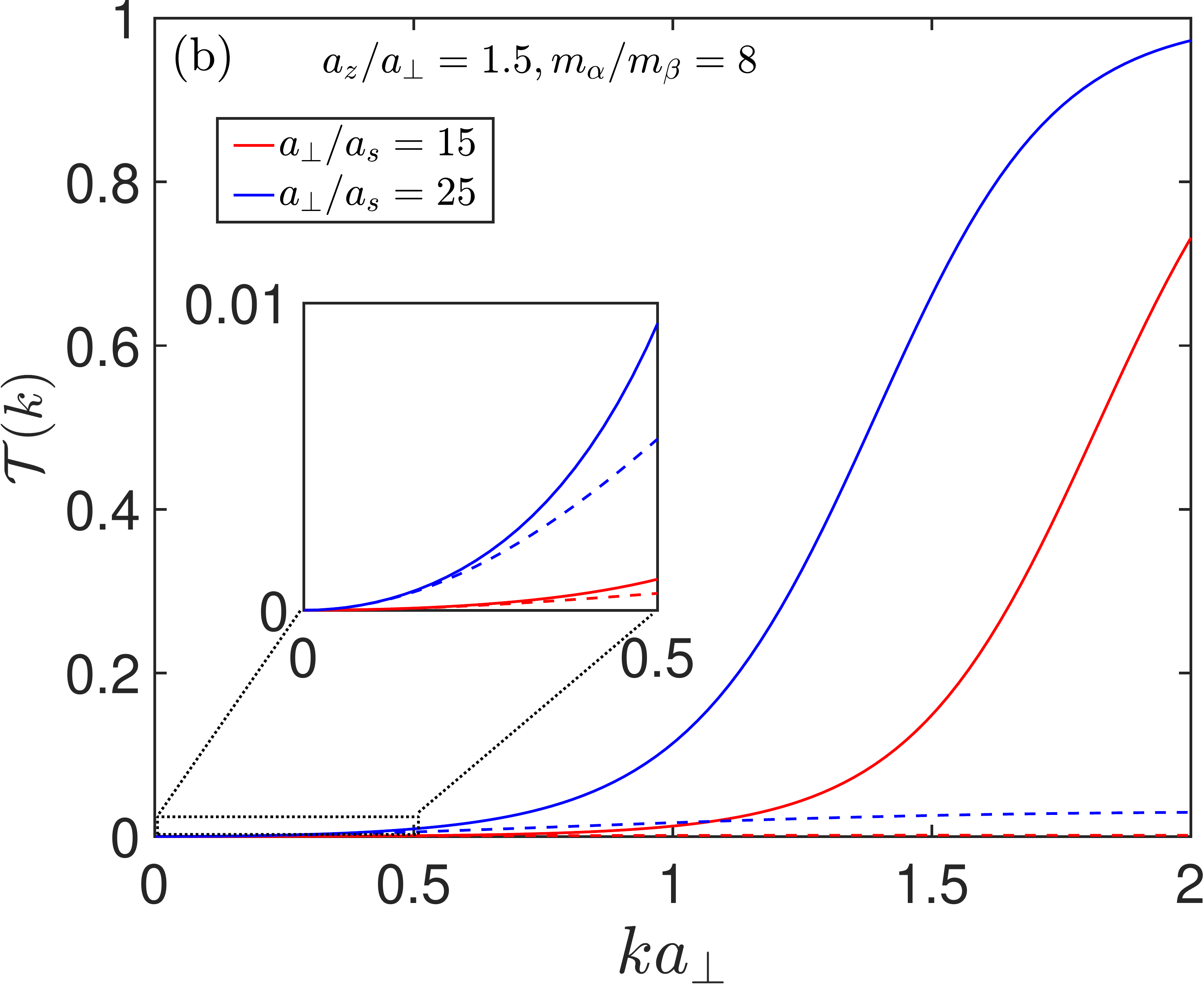}
\caption{The transmission coefficients ${\cal T}(k)$ as function of incident momentum $k$ of atom $\alpha$ for various cases with $a_s<0$ (a) and $a_s>0$ (b). Here we take $a_{z}/a_{\perp}=1.5$ and $m_{\alpha}/m_{\beta}=8$. The solid and dashed line are given by the calculations based on the exact quasi-(1+0)D Hamiltonian $H$ defined in Eq. (\ref{h}) and the effective 1D Hamiltonian $H^{\rm (eff)}$ defined in Eq. (\ref{heff2}), respectively.
\label{transmission}}
\end{figure}

Furthermore, 
when the confinement CLs $a_{\perp}$ and $a_z$ take some particular values, we may have
\begin{align}
a_{\rm e}=0,\nonumber
\end{align}
and thus the low-energy even-wave scattering amplitude $f^{\rm (e)}(k)$ can be maximally enhanced, i.e., $|f^{\rm (e)}(k)|\approx 1$. This is known as the even-wave CIR. Similarly, we may have $a_{\rm o}=\infty$ for some other particular values of $a_{\perp}$ and $a_z$. As a result, the odd-wave scattering amplitude $f^{\rm (o)}(k)$ is maximally enhanced ($|f^{\rm (o)}(k)|\approx 1$). This is known as the odd-wave CIR \cite{Girardeau}. It is clear that under an even--wave or odd-wave CIR the corresponding effective interaction intensity $g^{\rm(even)}$ or $g^{\rm(odd)}$ would be enhanced to infinity, respectively. In addition, when the system is in the region around a even- (odd-) wave CIR point, one can control $g^{\rm(even)}$ ($g^{\rm(odd)}$) by changing the confinement CLs $a_\perp$ and $a_z$. Since in our system the even and odd partial waves are decoupled, the appearance of even-wave CIR is actually independent of the odd-wave scattering, and vise versa. 

In this work we only consider  the even-wave CIRs.
We will show that when the mass ratio $m_\alpha/m_\beta$ is large enough, an even-wave CIR can occur even when $|a_s|$ is much smaller than the CLs $a_\perp$ and $a_z$ of the confinements of our system.

\subsection{The specific even-wave CIRs}

Now we study the even-wave CIRs of our system. To this end, we should calculate the even-wave scattering length $a_{\rm e}$ or the effective interaction intensity $g^{\rm (even)}$ defined in Eq. (\ref{geven}). We adopt the theoretical approach in our previous work where CIRs for cases with $m_\alpha=m_\beta$ are studied \cite{renKondo2}. The details of this method are presented in Appendix \ref{appendixA}.

In Fig.~\ref{fig2} we show $g^{\rm (even)}$ as a function of $a_\perp/a_s$, for the cases with different mass ratio $m_\alpha/m_\beta$, with $a_z/a_\perp=1$ (Fig.~\ref{fig2}(a-d)) and $a_z/a_\perp=1.5$ (Fig.~\ref{fig2}(e-h)). It is clearly shown that in each case multiple CIRs can appear for either $a_s>0$ or $a_s<0$. As pointed out in our previous work, that is due to the coupling between the relative and center-of-mass motion of the two atoms in the $z$-direction \cite{yvan,nishidamix,mixdimen_exp,yantingkondo,renKondo2}.  In Fig.~\ref{transmission} we further illustrate 
the transmission coefficient ${\cal T}(k)$, which is defined as
${\cal T}(k)=|1+f^{\rm (e)}(k)+f^{\rm (o)}(k)|^{2}$, for several cases. We compare the results given by  the quasi-(1+0)D calculation based on the explicit Hamiltonian $H$ in Eq. (\ref{h}) and the one from the effective 1D model $H^{\rm (eff)}$ of Eq. (\ref{heff2}).
It is clearly shown that in each case the
effective potential $V_{\rm eff}(z_\alpha)$ can reproduce the exact results in the low-energy limits, as mentioned in the above subsection.

\begin{figure}[h]
\centering
\includegraphics[width=0.35\textwidth]{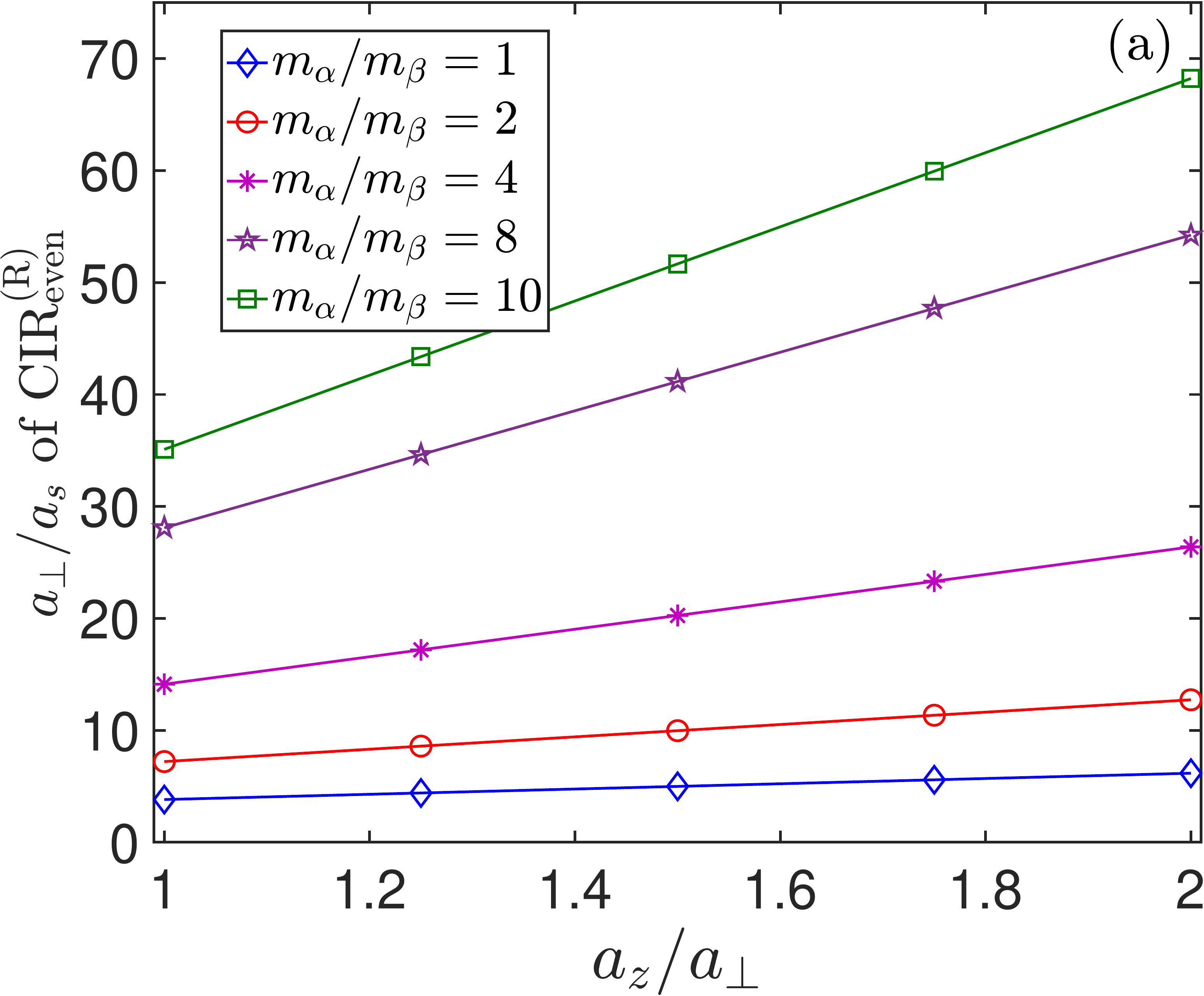}
\includegraphics[width=0.35\textwidth]{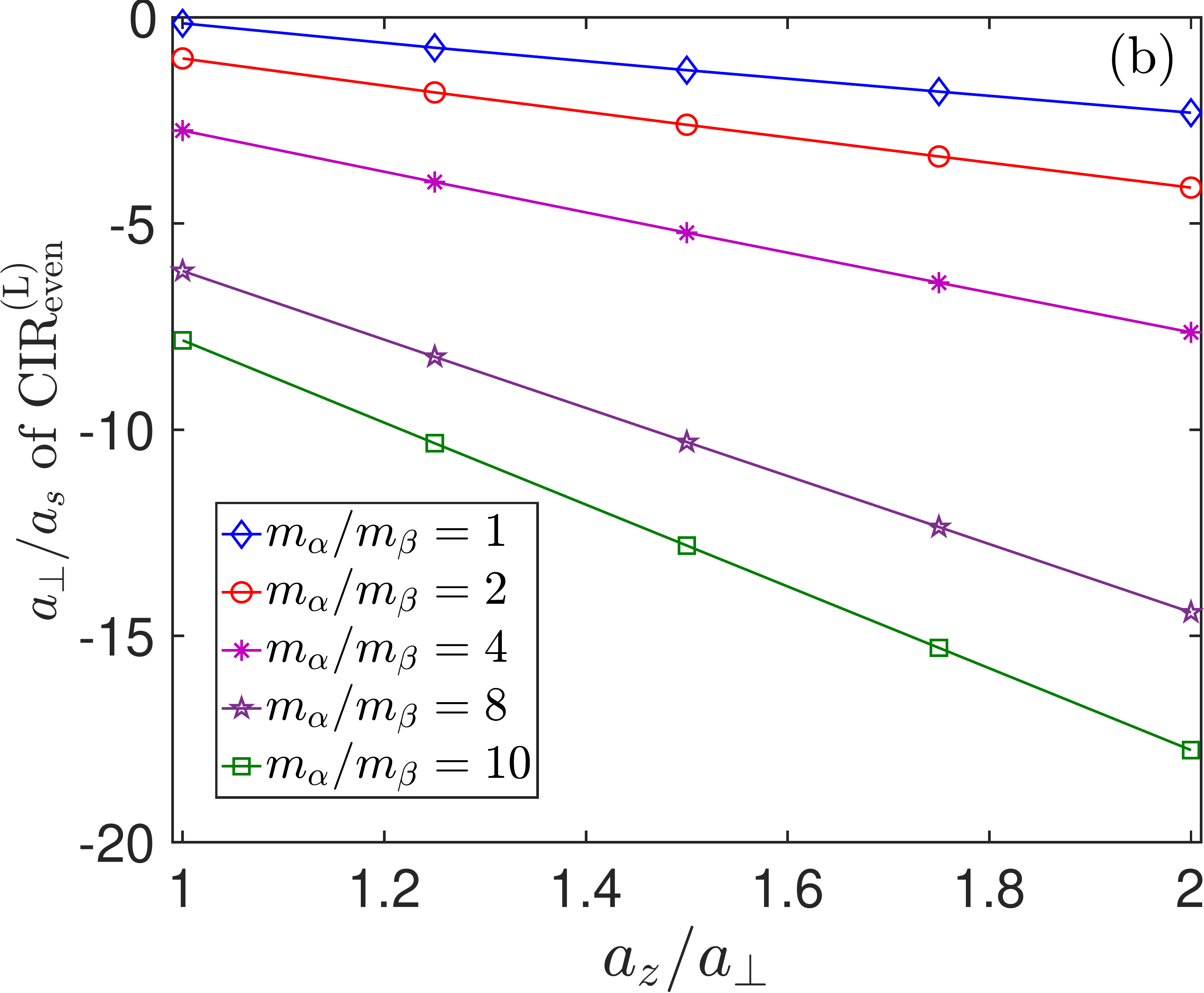}
\caption{(color online) Positions of CIR$_{\rm even}^{\rm (R)}$ and CIR$_{\rm even}^{\rm (L)}$ as functions of $a_{z}/a_{\perp}$ for different mass ratio.\label{fig3}}
\end{figure}

Furthermore, two broadest CIRs are located at the higher and lower ends of $a_\perp/a_s$, which are denoted as CIR$_{\rm even}^{\rm (R)}$ and CIR$_{\rm even}^{\rm (L)}$ in Fig.~\ref{fig2}, respectively. 
More importantly, as shown in Fig.~\ref{fig2}, when the mass ratio $m_\alpha/m_\beta$ is increased,
CIR$_{\rm even}^{\rm (R)}$ and CIR$_{\rm even}^{\rm (L)}$ are rapidly shifted to the places with larger $|a_\perp/a_s|$. 
For instance, when $a_z/a_\perp=1$ the CIR$_{\rm even}^{\rm (R)}$ occurs at $a_\perp=7.211a_s$ for $m_\alpha/m_\beta=2$, $a_\perp=14.12a_s$ for $m_\alpha/m_\beta=4$ and $a_\perp=28.09a_s$ for $m_\alpha/m_\beta=8$. Similarly, when $a_z/a_\perp=1$ the CIR$_{\rm even}^{\rm (L)}$ occurs at $a_\perp=-0.9989a_s$ for $m_\alpha/m_\beta=2$, $a_\perp=-2.747a_s$ for $m_\alpha/m_\beta=4$ and $a_\perp=-6.152a_s$ for $m_\alpha/m_\beta=8$. This is also shown in Fig.~\ref{fig3} where the positions of CIR$_{\rm even}^{\rm (L)}$ and CIR$_{\rm even}^{\rm (R)}$ are illustrated as functions of the confinement aspect ratio $a_{z}/a_{\perp}$ for different mass ratio $m_{\alpha}/m_{\beta}$. In addition, Fig.~\ref{fig3} also shows that for a fixed mass ratio, when $a_{z}/a_{\perp}$ is larger, CIR$_{\rm even}^{\rm (R)}$ and CIR$_{\rm even}^{\rm (L)}$ can appear for larger $|a_\perp/a_s|$.

According to our above results, when the mass ratio $m_\alpha/m_\beta$ of the two atoms is large enough, CIR$_{\rm even}^{\rm (R)}$ and CIR$_{\rm even}^{\rm (L)}$ can occur when the confinement CLs $a_z$ and $a_\perp$ are much larger than $|a_s|$ for $a_s>0$ and $a_s<0$, respectively. Namely the condition (i) in Sec. I can be satisfied by these two CIRs.

\begin{figure}[h]
\centering
\includegraphics[width=0.35\textwidth]{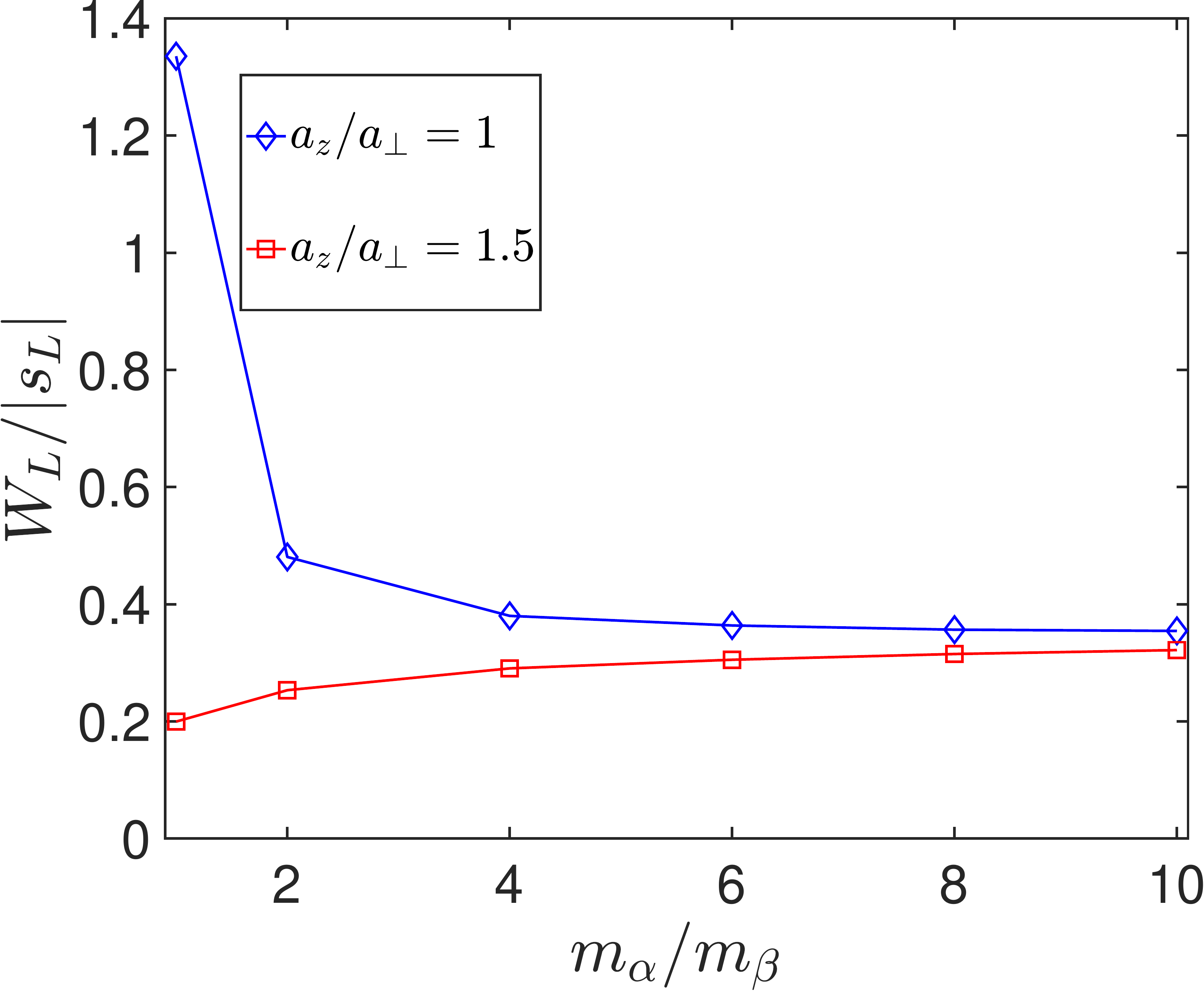}
\caption{(color online) $W_{L}/|s_{L}|$ of CIR$_{\rm even}^{\rm (L)}$ as a function of the mass ratio $m_{\alpha}/m_{\beta}$. As shown in the main text, $W_{L}$ is defined as the distance between the position of CIR$_{\rm even}^{\rm (L)}$ and the nearest zero-crossing point of $g^{\rm(even)}$ as illustrated in Fig.\ref{fig2}(d).
\label{width}}
\end{figure}

Now we consider the condition (ii)
in Sec. I, i.e.,
 the controllability of the effective interaction strength $g^{\rm (even)}$ in the region of CIR$_{\rm even}^{\rm (L,R)}$. As shown in Fig.~\ref{fig2}, using these CIRs one can control $g^{\rm (even)}$ via the confinement CLs $a_{\perp,z}$. This control is applicable when $g^{\rm (even)}$ is robust enough with respect to $a_{\perp,z}$, so that the fine-tuning of $a_{\perp,z}$ is not required. For our system we can analysis the robustness of $g^{\rm (even)}$ as follows. We first take the cases with $a_z/a_\perp=1$ as an example. As illustrated in Fig.~\ref{fig2} (d), we define the width $W_L$ of the CIR$_{\rm even}^{\rm (L)}$ as the distance between the position of this CIR and the nearest zero-crossing point of $g^{\rm(even)}$. For convenience, we further denote $s_L$ as the position of CIR$_{\rm even}^{\rm (L)}$ (i.e., CIR$_{\rm even}^{\rm (L)}$ appears when $a_\perp=a_z=s_La_s$). Thus, 
one can control $g^{\rm (even)}$ via tuning 
$a_{\perp,z}$ in the region between $(s_L-W_L)a_s$ and $(s_L+W_L)a_s$. Therefore, to realize a precise control for $g^{\rm (even)}$, the relative error of $a_{\perp,z}$ should be much less than $W_L/|s_L|$. In Fig.~\ref{width} we illustrate $W_L/|s_L|$
for various cases. It is shown that  we always have $W_L/|s_L|\gtrsim 30\%$. Therefore, the CIR$_{\rm even}^{\rm (L)}$ is applicable for the control of $g^{\rm (even)}$ when the relative error of $a_{\perp,z}$ is much less than $30\%$. This condition can be satisfied in most of the current experiments. 
Furthermore, the analysis for the cases with other values of $a_z/a_\perp$ as well as CIR$_{\rm even}^{\rm (R)}$ lead to a similar result. So we conclude that in the regions of CIR$_{\rm even}^{\rm (L)}$ and CIR$_{\rm even}^{\rm (R)}$ the effective interaction intensity $g^{\rm (even)}$ is robust enough with respect to $a_{\perp,z}$, and thus these CIRs can be used for the control of $g^{\rm (even)}$.

On the other hand, we also notice that the CLs of the laser trapping potentials are proportional to the quadratic root of the corresponding laser intensities, i.e., $a_{\perp,z}\propto I_{\perp,z}^{1/4}$, with $I_\perp$ ($I_z$) being the intensity of the laser beam which creates the transverse (axial) confinement. In many experiments, the intensities of the trapping lasers can be varied by a factor of at most 3 or 4, which yields that $a_{\perp,z}$ can be tuned by a factor of at most $1.3-1.4$. The variation of $a_{\perp,z}$ in a broader range may be difficult. As a result, for a certain system one can tune $g^{\rm (even)}$ in a finite region with the help of CIR$_{\rm even}^{\rm (L,R)}$ (e.g., the strongly-interacting region with large $|g^{\rm (even)}|$), but it is not easy to freely control $g^{\rm (even)}$ in the whole range from $-\infty$ to $+\infty$.


\section{Analysis based on the BOA}

In the above section, we show that for our quasi-(1+0)D system CIRs can appear when the 3D scattering length is much smaller than the confinement CLs. In this section, we qualitatively explain the appearance of these specific CIRs with an analysis based on the BOA.

As shown above, for our system the atom $\alpha$ is much heavier than the atom $\beta$, i.e., $m_\alpha\gg m_\beta$. Therefore, the mass of atom $\alpha$ is also much larger than the reduced mass $\mu$ of the two atoms, which is the effective mass of the two-atom relative motion. Consequently, both of the coordinate $z_\beta$ of the atom $\beta$ and the transverse coordinate ${\bm \rho}$ of the two-atom relative motion are the ``fast variable" of our system, while the coordinate $z_\alpha$ of atom $\alpha$ is the only ``slow variable". Therefore, in the spirit of BOA, the axial motion of atom $\alpha$ is governed by a 1D Hamiltonian
\begin{align}
H_{\rm BOA}=-\frac{\hbar^{2}}{2m_{\alpha}}\frac{\partial^{2}}{\partial z_{\alpha}^{2}}+V_{\rm BOA}(z_\alpha).\label{heff}
\end{align}
Here the potential $V_{\rm BOA}(z_\alpha)$ is the energy of fast variables for a fixed axial position $z_\alpha$ of the atom $\alpha$, i.e., the ground-state energy of the Hamiltonian $H_\perp+H_z^{\beta}+V$ with $z_\alpha$ being an classical parameter (c-number). Furthermore, since the 3D scattering length $a_s$ is very small, i.e., the bare interaction $V({\bf r})$ in Eq.(\ref{HY}) is very weak, we can treat $V({\bf r})$ as a first-order perturbation in the calculation of $V_{\rm BOA}(z_\alpha)$. 
A straightforward calculation yields
\begin{align}
V_{\rm BOA}(z_\alpha)&=\left(\frac{a_s}\mu\right)\frac{2\hbar^2}{\sqrt{\pi} a_\perp^2 a_z}e^{-z_\alpha^2/a_z^2}.\label{veff2}
\end{align}

We first consider the case with $a_s<0$. Eq. (\ref{veff2}) shows that in this case $V_{\rm BOA}(z_\alpha)$ is always negative, i.e., it is a 1D finite-range potential well which is determined by the confinement CLs $\{a_z, a_\perp\}$, as well as the factor $a_{s}/\mu$. Therefore, if the reduced mass $\mu$ is sufficiently small (i.e., the mass ratio $m_\alpha/m_\beta$ is large enough), 
the potential well $V_{\rm BOA}(z_\alpha)$ would be deep enough and thus the 1D even-wave scattering resonance can appear (i.e., the even-wave scattering length can become zero). Thus, as shown in Fig.~\ref{fig2} and Fig.~\ref{fig3}(b), when $m_\alpha/m_\beta$ becomes larger, the CIRs can occur for smaller $a_{s}$. On the other hand, when the ratio $a_{s}/\mu$ is fixed, the shape of $V_{\rm BOA}(z_\alpha)$ still changes with $a_\perp$ and $a_z$. Resonance can be induced when these two parameters are tuned to some particular values.
 
\begin{figure}[t]
\centering
\includegraphics[width=0.35\textwidth]{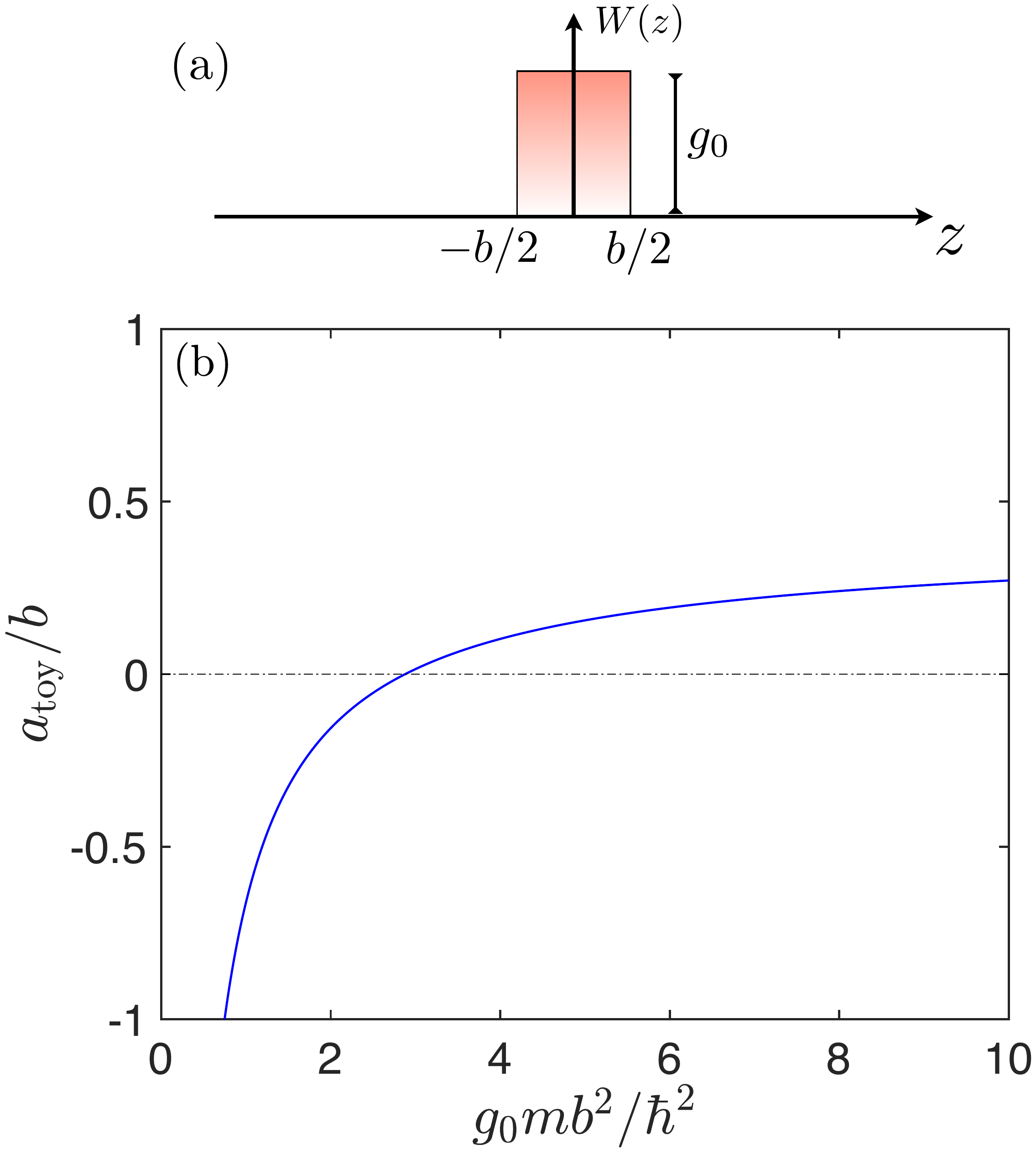}
\caption{(color online) (a) The pure-1D square potential barrier $W(z)$ in the Hamiltonian (\ref{sq}). (b) Scattering length $a_{\rm toy}$ of the pure-1D square barrier as function of width $b$ and height $g_{0}$. An even-wave CIR occurs when $g_0\approx 2.9\hbar^2/(mb^2)$.\label{fig4}}
\end{figure}

For the case with $a_s>0$, the potential $V_{\rm BOA}(z_\alpha)$ given by Eq. (\ref{veff2}) is always positive. I.e., $V_{\rm BOA}(z_\alpha)$ is a potential barrier. In this case the appearance of the resonance seems to be counter-intuitive. Nevertheless, for 1D scattering a zero-energy even-wave resonance really can occur for a potential barrier which is always positive. As an example, we consider the scattering of a particle on a square potential barrier in the $z$-axis, as shown in Fig.~\ref{fig4}(a). The Hamiltonian of this toy model is 
\begin{align}
h_{\rm toy}=-\frac{\hbar^{2}}{2m}\frac{\partial^{2}}{\partial z^{2}}+W(z),\label{sq}
\end{align}
with $W(z)=g_0$ for $|z|\leq b/2$ and $W(z)=0$ for $|z|>b/2$ ($b>0$) (Fig.~\ref{fig4}(a)). A direct calculation shows that the even-wave scattering length $a_{\rm toy}$ for this toy model is
\begin{align}
a_{\rm toy}=\frac{b}{2}-\frac{\coth(b\sqrt{2mg_{0}/\hbar^{2}}/2)}{\sqrt{2mg_{0}/\hbar^{2}}}. \label{atoy}
\end{align}
In Fig.~\ref{fig4}(b), $a_{\rm toy}$ is shown as a function of the width $b$ and the height $g_0$ of the potential. Eq. (\ref{atoy}) yields that we have $a_{\rm toy}=0$ when $g_0\approx 2.9\hbar^2/(mb^2)$. An even-wave resonance can occur under this condition. 

Therefore, similar as in the case of $a_s<0$, when $a_s>0$ an even-wave CIR can appear
provided that the mass ratio $m_\alpha/m_\beta$ is so large that the strength $a_s/\mu$ of $V_{\rm BOA}(z_\alpha)$ is strong enough, and the confinement CLs $a_z$ and $a_\perp$ are tuned to some particular values.


To conclude, the above analysis shows that the specific CIRs of our system mainly results from the following two facts. First, the potential $V_{\rm BOA}(z_\alpha)$ depends on the ratio $a_{s}/\mu$, which is essentially because the Huang-Yang pseudo potential $V$ is proportional to $a_{s}/\mu$. Second, an even-wave resonance can always occur for a finite-range 1D potential, no matter if it is a potential well or a potential barrier.

\section{Summary}

In this work we study the low-energy scattering between a heavy atom $\alpha$ moving in quasi-1D confinement and a localized light atom $\beta$. We show that 
if the atom $\beta$ is light enough, two specific CIRs can appear even if the 3D inter-atomic scattering length $a_s$ is much smaller than the CLs of the confinements, for either $a_s>0$ or $a_s<0$. With these two CIRs, one can realize strong effective inter-atomic interaction even if the 3D background interaction between these two atoms is very weak, and thus realize a strongly-interacting impurity system without the help of MFR. So far the mixture of ultracold atoms with different mass ratio has been realized by many experimental groups, e.g, the mixtures of $^{173(4)}$Yb- $^{7}$Li\cite{Yb-Li}, $^{161}$Dy- $^{40}$K \cite{Dy-K} and $^{40}$K-$^{6}$Li \cite{K-Li1,K-Li2,K-Li3}. Our proposal is very hopeful to be realized in these experimental systems.

We plan in the future to find this kind of specific CIRs, which can appear for small $a_s$ and is broad enough for experimental control, in more types of confinements. As shown in Sec. I, they could be very useful for quantum simulations and precision measurements, especially the ones which are based on spin-dependent inter-atomic interaction and should be performed at a very low magnetic field or $B=0$.

\begin{acknowledgments}
We thank the referees for improving the quality of this manuscript.
This work is supported by the National Key R$\&$D Program of China (Grant No. 2018YFA0307601 (RZ), 2018YFA0306502 (PZ)), NSFC Grant No. 11804268 (RZ), 11434011(PZ), 11674393(PZ), as well as the Research Funds of Renmin University of China
under Grant No. 16XNLQ03(PZ).
\end{acknowledgments}

\appendix
\begin{widetext}
\section{Calculation of Even-Wave Scattering Length for Quasi-(1+0)D System}
\label{appendixA}
In this appendix, we present the detailed calculation of the even-wave scattering length $a_{\rm e}$ defined in Eq.(\ref{fe}) of the main text. As mentioned in our main text, our approach is the same as our previous work Ref. \cite{renKondo2}, where we performed this calculation for the equal-mass case with $m_\alpha=m_\beta$. Since the calculation method is introduced in detail in Ref. \cite{renKondo2}, here we only show the basic idea of this method and the formulas which are different from the ones in the equal-mass case.

As shown in Ref. \cite{renKondo2}, the scattering wave function $\Psi^{(+)}({\bm \rho},z_\alpha,z_\beta)$ introduced in Sec. II. A satisfies the Lippmann-Schwinger type equation 
\begin{align}
\Psi^{(+)}({\bm \rho},z_{\alpha},z_{\beta})=\Psi^{(0)}({\bm \rho},z_{\alpha},z_{\beta})+\frac{2\pi\hbar^{2} a_{s}}{\mu}\int dz'G_{E}({\bm \rho},z_{\alpha},z_{\beta};{\bf 0},z',z')\eta(z'),\label{lse}
\end{align}
where $\Psi^{(0)}({\bm \rho},z_{\alpha},z_{\alpha})$ is the incident state defined in Eq.(\ref{incoming-state}) of the main text and $E=\frac{\hbar^{2}k^{2}}{2m_{\alpha}}+\frac{\hbar\omega_{z}}{2}+\hbar\omega_{\perp}$ is the scattering energy. Here $G_{E}({\bm \rho},z_{\alpha},z_{\beta};{\bf 0},z_{\alpha}',z_{\beta}')$ is the Green's function defined as 
\begin{align}
\label{G0}
G_{E}({\bm \rho},z_{\alpha},z_{\beta};{\bm \rho}',z_{\alpha}',z_{\beta}')=&\langle{\bm \rho},z_{\alpha},z_{\beta}|\frac{1}{E+i0^{+}-H_{0}}|{\bm \rho}',z_{\alpha}',z_{\beta}'\rangle
\end{align}
and $\eta(z)$ is the regularized scattering wave function 
\begin{align}
\eta(z)=\left.\frac{\partial}{\partial z_{r}}\left[z_{r}\psi^{(+)}\left({\bf 0},z+\frac{m_{\beta}}{M}z_{r},z-\frac{m_{\alpha}}{M}z_{r}\right)\right]\right|_{z_{r}\rightarrow0}\label{eta}
\end{align}
with $M=m_{\alpha}+m_{\beta}$. Upon substituting Eq.(\ref{lse}) into Eq.(\ref{eta}), one immediately obtains the integral equation of $\eta(z)$
\begin{align}
\eta(z)=\Psi^{(0)}({\bf 0},z,z)+\frac{2\pi\hbar^{2} a_{s}}{\mu}\left.\frac{\partial}{\partial z_{r}}\left[z_{r}\int dz'G_{E}\left({\bf 0},z+\frac{m_{\beta}}{M}z_{r},z-\frac{m_{\alpha}}{M}z_{r};{\bf 0},z',z'\right)\eta(z')\right]\right|_{z_{r}\rightarrow0^{+}}.\label{eta2}
\end{align}
Solving Eq.(\ref{eta2}), one could obtain $\eta(z)$ and hence the scattering state $\Psi^{(+)}({\bm \rho},z_{\alpha},z_{\beta})$. Subjecting to the out-going boundary condition in Eq.(\ref{outgoing}), one could find the even-wave scattering amplitude as follows
\begin{align}
\label{fe1}
f^{(\rm e)}(k)=&\frac{m_{\alpha}}{\mu}\frac{(2\pi)^{3/2}a_{s}}{ik}\sqrt{\frac{\mu\omega_{\perp}}{\hbar\pi}}\int dz'\cos(ikz')\phi_{0}^{*}(z')\eta(z').
\end{align}
In the zero momentum limit, the even-wave scattering amplitude can be written as
\begin{align}
f^{\rm (e)}(k)&\approx\frac{-1}{1+ika_{\rm e}};\label{fe2}
\end{align}


Therefore, if we can obtain the function $
\eta(z)$, we can obtain the even-wave scattering length $a_{\rm e}$ via Eq.(\ref{fe1}) and Eq.(\ref{fe2}). To derive $\eta(z)$, we re-write Eq. (\ref{eta2}) as an integral equation for $\eta(z)$. As in Ref. \cite{renKondo2}, we first re-express the Green's function as
\begin{align}
G_{E}({\bf 0},z_{\alpha},z_{\beta};{\bf 0},z',z') = \frac{\mu\omega_{\perp}}{\hbar\pi}g(z_{\alpha},z_{\beta};z',z')+G_{E^{\prime}}({\bf 0},z_{\alpha},z_{\beta};{\bf 0},z',z')\label{g03}
\end{align}
Here the two terms are given by
\begin{align}
g(z_{\alpha},z_{\beta};z',z')=-i\frac{m_{\alpha}}{\hbar^{2}}\frac{e^{ik|z_{\alpha}-z'|}}{k}\phi_{0}(z_{\beta})\phi_{0}^{*}(z')-\frac{m_{\alpha}}{\hbar^{2}}\sum_{n_{z}=1}^{\infty}\frac{e^{-\sqrt{2m_{\alpha}n_{z}\omega_{z}/\hbar-k^{2}}|z_{\alpha}-z'|}}{\sqrt{2m_{\alpha}n_{z}\omega_{z}/\hbar-k^{2}}}\phi_{n_{z}}(z_{\beta})\phi_{n_{z}}^{*}(z')\label{littleg0}
\end{align}
and 
\begin{align}
G_{E^{\prime}}\left({\bf 0},z_{\alpha},z_{\beta};{\bf 0},z^{\prime},z^{\prime}\right)=-\int_{0}^{\infty}d\beta e^{\beta E^{\prime}}K_{\beta}\left({\bf 0},z_{\alpha},z_{\beta};{\bf 0},z^{\prime},z^{\prime}\right)\label{gep2}
\end{align}
with $E^{\prime}=E-2\hbar\omega_{\perp}$ which is smaller than the threshold energy of $H_{0}$, and
\begin{align}
{\cal K}_{\beta}\left({\bf 0},z_{\alpha},z_{\beta};{\bf 0},z^{\prime},z^{\prime}\right) =& \frac{\mu\omega_{\perp}}{2\pi\hbar\sinh(\hbar\omega_{\perp}\beta)}\times\sqrt{\frac{m_{\alpha}}{2\pi\hbar\beta}}\exp\left[-\frac{m_{\alpha}(z_{\alpha}-z^{\prime})^{2}}{2\hbar\beta}\right]\nonumber \\
 &\times\sqrt{\frac{m_{\beta}\omega_{z}}{2\pi\hbar\sinh(\hbar\omega_{z}\beta)}} \exp\left[-\frac{m_{\beta}\omega_{z}\left[\left( z_{\beta}^{2}+z^{\prime2}\right) \cosh(\hbar\omega_{z}\beta)-2z_{\beta}z^{\prime}\right]}{2\hbar\sinh(\hbar\omega_{z}\beta)}\right].\label{calkb}
\end{align}


Using the result Eq. (\ref{g03}) and the approach present in Ref. \cite{renKondo2}, we 
find that Eq. (\ref{eta2}) can be converted to an integral equation for $\eta(z)$
\begin{align}
\eta(z) = &\Psi^{(0)}({\bf 0},z,z)+2\hbar\omega_{\perp}a_{s}\int dz^{\prime}g(z,z;z',z')\eta(z')+\frac{2\pi\hbar^{2} a_{s}}{\mu}\left\{F_{1}(z)\eta(z)+\int dz'F_{2}(z,z^{\prime})\left[\eta(z')-\eta(z)\right]\right\},\label{eta4}
\end{align}
which is solvable numerically. Here 
\begin{align}
F_{1}(z)= & -\frac{\mu^{3/2}}{(2\pi\hbar^{2})^{3/2}}\int_{0}^{\infty}d\beta\left[\frac{\hbar\omega_{\perp}\sqrt{(m_{\alpha}+m_{\beta})\omega_{z}}\exp\left(\beta E^{\prime}-\frac{m_{\beta}\omega_{z}\left[m_{\beta}\omega_{z}\hbar\beta+2m_{\alpha}\tanh(\hbar\omega_{z}\beta/2)\right]}{2\hbar[m_{\alpha}+m_{\beta}\omega_{z}\hbar\beta\coth(\hbar\omega_{z}\beta)]}z^{2}\right)}{\sinh(\hbar\beta\omega_{\perp})\sqrt{m_{\beta}\omega_{z}\beta\cosh(\hbar\omega_{z}\beta)+m_{\alpha}\sinh(\hbar\omega_{z}\beta)/\hbar}}-\frac{1}{\beta^{3/2}}\right]
\label{f1}
\end{align}
and 
\begin{align}
\label{f2}
F_{2}(z,z^{\prime})=-\int_{0}^{\infty}d\beta e^{\beta E^{\prime}}{\cal K}_{\beta}\left({\bf 0},z,z;{\bf 0},z^{\prime},z^{\prime}\right)
\end{align}

We numerically solve Eq. (\ref{eta4}) and derive the function $
\eta(z)$. Substituting the result into Eq.(\ref{fe1}) and Eq.(\ref{fe2}), we finally obtain the even-wave scattering length $a_{\rm e}$.

\end{widetext}

%
%
%

\bibliography{references}

\end{document}